\documentclass[usenatbib]{mn2e}

\usepackage{graphicx}
\usepackage{amsmath}
\usepackage{amssymb}

\def\thee{\theta_E}
\def\tlens{t_{\rm lens}}
\def\amp{A}

\def\up#1{\hbox{\raise.5ex\hbox{#1}}}

\begin{document}

\title{Microlensing masses via photon bunching}

\author[P. Saha]{Prasenjit Saha\\Physik-Institut, University of
  Zurich, Winterthurerstrasse 190, 8057 Zurich, Switzerland}

\maketitle

\begin{abstract}
In microlensing of a Galactic star by a brown dwarf or other compact
object, the amplified image really consists of two unresolved images
with slightly different light-travel times.  The difference (of order
a microsecond) is $GM/c^3$ times a dimensionless factor depending on
the total magnification.  Since magnification is well-measured in
microlensing events, a single time-delay measurement would provide the
mass of the lens, without degeneracies.  The challenge is to find an
observable that varies on sub-microsecond time scales.
This paper notes that the narrow-band intensity of the unresolved
image pair will show photon bunching (the Hanbury Brown and Twiss
effect), and argues that the lensed intensity will have an
auto-correlation peak at the lensing time delay.  The ultrafast
photon-counting technology needed for this type of measurement exists,
but the photon numbers required to give sufficient signal-to-noise
appear infeasible at present.  Preliminary estimates suggest
time-delayed photon bunching may be measurable for lensed early-type
main-sequence stars at $\sim10\rm\,kpc$, with the help of 30\thinspace
m-class telescopes.
\end{abstract}


\section{Introduction}

In Galactic microlensing there are two lensed images (more if the lens
is binary), but there is no prospect of resolving them.  Everything
has to be inferred from a point image.  

The principal observable is (i)~the time-dependent brightness
amplification (or light curve).  \cite{1936LinkF,1937BuAst..10...73L}
and independently \cite{1936Sci....84..506E} discussed it long before
it became feasible to observe, with strikingly contrasting views on
whether it could ever be observed.  Later but still in the dream-time
of gravitational lensing, \cite{1966MNRAS.134..315R} drew attention to
two more observables: (ii)~parallax, which in the context of
microlensing refers to the dependence of the light curve on observer
location; and (iii)~apparent proper motion of the image.  Parallax
depends on the spatial scale of the lens, while proper motion reveals
its angular scale.  Combining parallax and proper motion with the
light curve yields the lens mass and distance.  For parallax, Refsdal
envisioned a spacecraft elsewhere in the solar system, an idea revived
more recently \cite[e.g.,][]{1996ApJ...462..705B}.  For proper
motions, Refsdal assumed that both source and lens would be visible
stars. An alternative proposal, for when the lens is dark, is to
monitor the lensed-image centroid for proper motion
\citep[e.g.,][]{1998ApJ...502..538B}.
\cite{1995lssu.conf..326V,1998MNRAS.294..747V} predicted a further
observable: (iv)~chromatic amplification, meaning colour, spectral and
polarisation changes in the image due to variations of these across
the face of the source star.  Given a good model for the stellar
atmosphere, chromatic amplification becomes in effect a surrogate for
proper motion.  Various observables come together beautifully in
\cite{2009ApJ...698L.147G}, where chromatic amplification gives the
angular scale, while the parallax is large enough to measure from
different ground locations, and combined with the main light curve
they supply the lens mass.  But this cases are exceptional; as a rule,
mass measurement in microlensing involves parameter degeneracies.

This paper will consider yet another possible observable in an
unresolved microlensed image.  This is the lensing time delay, or the
difference in light travel time between the two lensed images.  The
formula for it is derived in Section~\ref{sec:arriv} and comes to
\begin{equation} \label{eqn:tdel}
\Delta\tlens = \frac{2GM}{c^3} \left( x^2 - 1/x^2 + 4\ln x \right)
\end{equation}
where
\begin{equation} \label{eqn:xmag}
x^4 = \frac{\amp+1}{\amp-1}
\end{equation}
and $\amp$ is the brightness amplification or total magnification.
Since $\amp$ at any stage of a microlensing event is known accurately
from the light curve, a single measurement of $\Delta\tlens$ would
automatically measure the mass.

Time-delay measurements are an active area for galaxy and cluster
lenses at cosmological distances \citep[e.g.][]{2016ApJ...820...50R}.
There is also research on modelling the mass distributions of the
lenses (as the point-mass approximation is not applicable to galaxies
and cluster lenses) to infer cosmological parameters
\citep[e.g.,][]{2014MNRAS.437..600S} or mass substructures
\citep{2015PASJ...67...21M}.  And long before the first time-delay
measurements, \cite{1964MNRAS.128..307R,1966MNRAS.132..101R} was
already advocating exploiting time-varying sources to measure time
delays from galaxy lenses in order to measure cosmological parameters.
Why then did he not propose time-delay measurements in his
microlensing paper \citep{1966MNRAS.134..315R}?  We can guess the
reason: the scale of $\Delta\tlens$ is utterly different ---
microseconds for brown-dwarf lenses versus days to years for galaxy or
cluster lenses --- and stars are not known to have intrinsic
brightness variations on sub-microsecond scales.  Measuring
microlensing time delays seems hopeless.

Yet perhaps not.  Section~\ref{sec:hbt} below will argue that
fluctuations inherent in incoherent light (wave noise) could be used
to advantage. The idea is to measure the brightness fluctuations, at
any one stage of the microlensing event, with nanosecond time
resolution.  The photon statistics will then not be quite Poisson, but
will show correlations.  In particular, the auto-correlation of the
photon arrival times will show the following features.
\begin{itemize}
\item First, there will be a peak at zero.  This is the well-known
  phenomenon of photon bunching.
\item Also expected --- and this is the main prediction of this paper
  --- are two smaller peaks at $\pm\Delta\tlens$, corresponding to the
  lensing time delay, or the difference in light travel time between
  the two lensed images.
\end{itemize}

Section~\ref{sec:snr} estimates the number of photons that would be
needed to measure the secondary peaks in the intensity
auto-correlation, assuming the effect is present.  If nearby bright
stars were microlensed, the auto-correlation peaks would be easy to
measure.  For stars at $\sim10\rm\,kpc$, which is where most
microlensed sources are, the necessary signal-to-noise appears
unachievable at present.  But upcoming developments (30\thinspace m
telescopes, sub-nanosecond photon-counting arrays) could make the
effect accessible for the brightest microlensed events.

\section{Microlensing time delays}\label{sec:arriv}

To derive the expression (\ref{eqn:tdel}) for the time delay in
microlensing, let us consider the arrival-time surface \citep[see
  e.g.,][]{1986ApJ...310..568B}.  The arrival time, up to an additive
constant, of a virtual photon coming from the direction $\theta$ but
having originated at a source in the direction $\beta$ is
\begin{equation}
ct = \frac{D_LD_S}{2D_{LS}} (\beta-\theta)^2 - \frac{4GM}{c^2} \ln\theta
\end{equation}
where $D_L,D_S,D_{LS}$ are respectively the distances to the lens,
source, and from lens to source.  In terms of the angular Einstein
radius
\begin{equation}
\thee^2 \equiv \frac{4GM}{c^2} \frac{D_{LS}}{D_SD_L}
\end{equation}
the arrival time takes a cleaner form:
\begin{equation} \label{eqn:arriv}
ct = \frac{4GM}{c^2}
    \left( \frac{(\beta-\theta)^2}{2\thee^2} - \ln\theta \right) \,.
\end{equation}
Actual photons take paths for which the arrival time is stationary, or
$dt/d\theta=0$.  This condition gives the usual lens equation
\begin{equation}
\beta = \theta - {\thee^2}/\theta \,.
\end{equation}
Solutions of the lens equation can be conveniently written in terms of
a new variable $x$ as
\begin{equation} \label{eqn:impos}
\theta = x\,\thee \ \hbox{and}\ \theta = -\thee/x \,.
\end{equation}
The source position corresponding to both images is
\begin{equation} \label{eqn:spos}
\beta=(x-1/x)\,\thee \,.
\end{equation}
Substituting the image and source positions
(\ref{eqn:impos},\ref{eqn:spos}) into the arrival time
(\ref{eqn:arriv}) gives the value for each image.  The difference
between the arrival times of the two images then simplifies to the
expression (\ref{eqn:tdel}) for the time delay.

In order to relate $x$ to the brightness amplification, recall the
well-known expression
\begin{equation}
\left(1 - \frac{\thee^4}{\theta^4}\right)^{-1}
\end{equation}
for the magnification due to a point lens.  The absolute magnification
at the two images comes to
\begin{equation}
\amp_1 = \frac{x^4}{x^4-1} \ \hbox{and}\ \amp_2 = \frac1{x^4-1}
\end{equation}
and the sum of these relates $x$ to the total magnification
(Eq.~\ref{eqn:xmag}).

\begin{figure}
\centering
\includegraphics[width=.973\hsize]{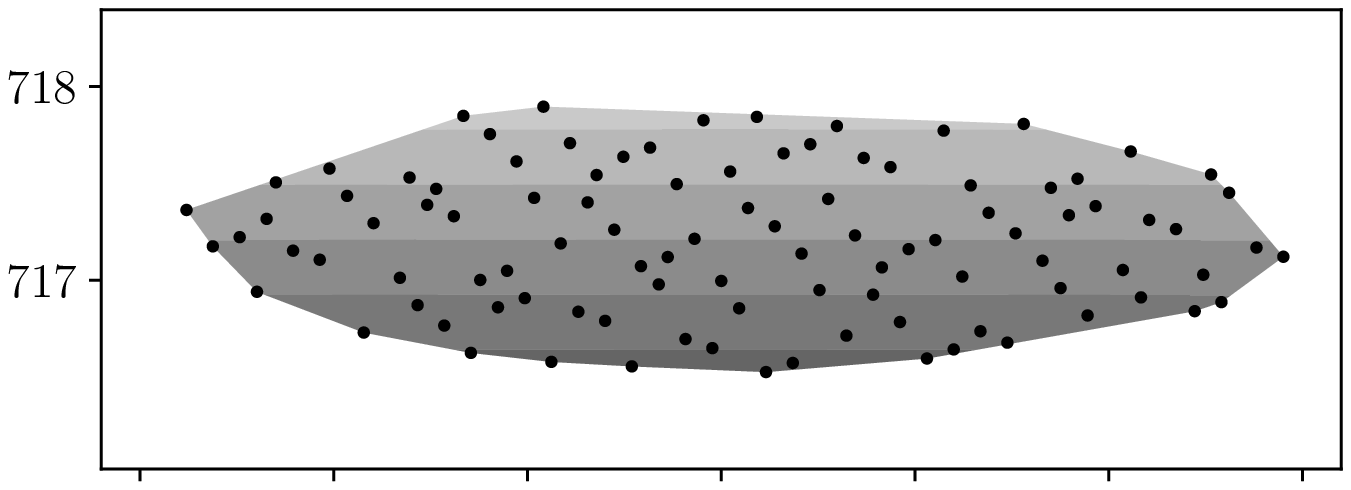}\kern-6pt\\
\includegraphics[width=.973\hsize]{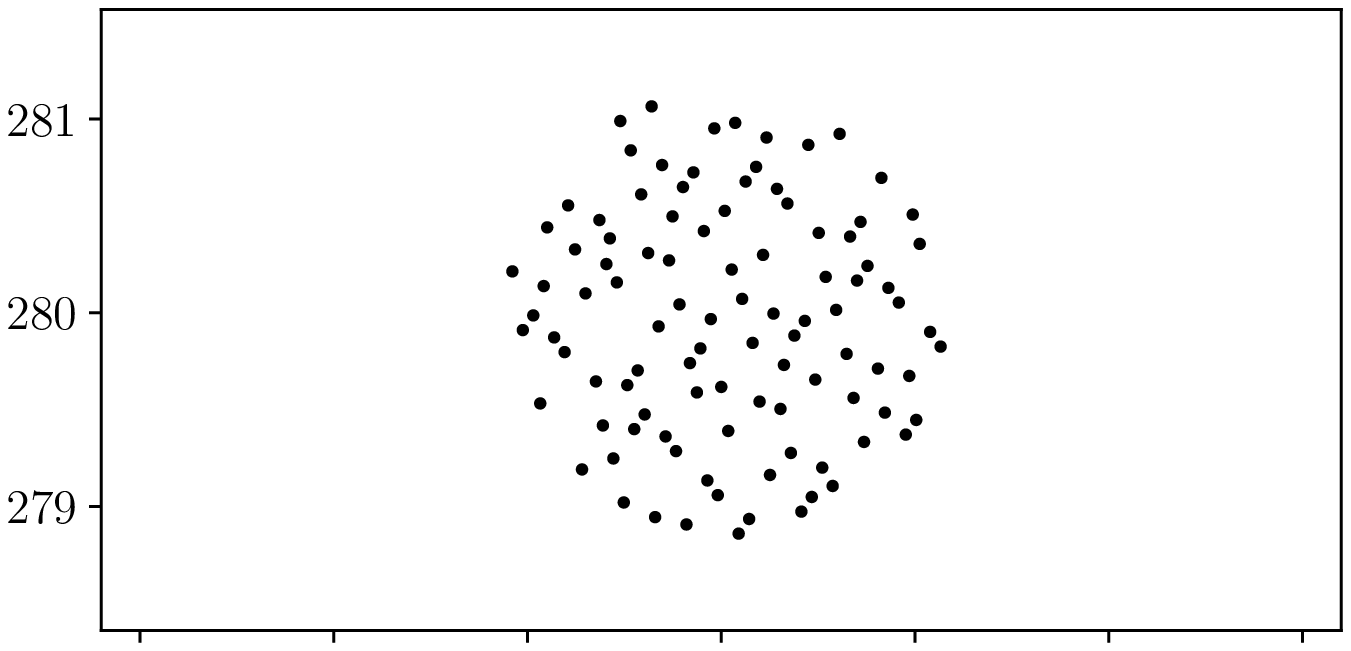}\kern-6pt\\
\includegraphics[width=\hsize]{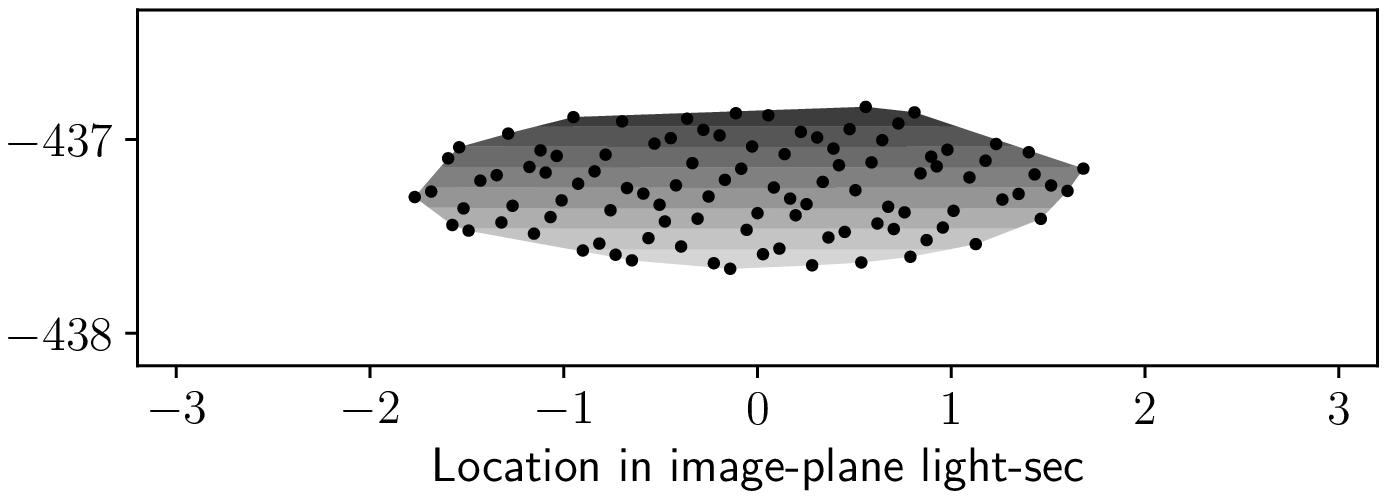}
\caption{\label{fig:zones} A representation of the lensed images and
  arrival times for a solar-sized source in the lensing configuration
  (\ref{eqn:params}).  The middle panel shows the source while the
  other two panels show the lensed images.  (The lens is at the
  origin.)  The dots represent a quasi-random sample of 100 zones on
  the source.  The grey-scale indicates the spread of arrival times
  within each image (darker for later arrival).  The spread is
  $\Delta\Delta\tlens=5\rm\,ns$ for the earlier image, and $8\rm\,ns$
  for the lower image.  The systemic difference between the images is
  much more, with the lower image arriving $\Delta\tlens=1584\rm\,ns$
  later on average.}
\end{figure}

The preceding applies only to a point source lensed by a point lens.
For a realistic source, even an unresolved source, it is necessary to
consider the effect of finite size.  From the expressions
(\ref{eqn:tdel}) for the time delay and (\ref{eqn:spos}) for the
source position, it follows that
\begin{equation}
\frac{d}{d\beta} \Delta\tlens = \frac{4GM}{c^3} \frac{x + 1/x}{\thee}
\end{equation}
indicating that
\begin{equation}
\frac{\Delta\Delta\tlens}{\Delta\tlens} \sim \frac{\Delta\beta}{\thee}
\end{equation}
where $\Delta\Delta\tlens$ stands for the spread in time delays across
the source.  As an example, consider the configuration
\begin{equation}\label{eqn:params}
\begin{aligned}
&D_L = 4\,{\rm kpc} \quad &D_S = 8\,{\rm kpc} \\
&M = 0.08M_\odot &\beta = {\textstyle\frac12} \thee
\end{aligned}
\end{equation}
which would be typical of microlensing events. The projected Einstein
radius will be $D_L\thee=1.1\rm\,au$ at the lens, and
$D_S\thee=2.2\rm\,au$ at the source.  If the source is Sun-sized, it
will be much smaller than the projected Einstein radius, and the
point-lens approximation is reasonable.  Furthermore,
$\Delta\Delta\tlens\ll\Delta\tlens$.  The spread in time delays will,
however, be orders of magnitude larger than the $1/\nu$ of light.
This is the reason lensed images do not produce interference fringes
on the ground \citep[cf.][]{1964MNRAS.128..295R,1996IAUS..173..407P}.
Figure~\ref{fig:zones} shows a numerical computation of arrival times
from an extended source. We see that for a Sun-sized source in the
lensing configuration (\ref{eqn:params}),
$\Delta\tlens\simeq1.6\times10^{-6}\rm\,s$ and
$\Delta\Delta\tlens\sim10^{-8}\rm\,s$.

We conclude that microlensing time delays could be measured if the
light from the source has fluctuations that are faster than the
microsecond scale, but not so fast that they average out within ten
nanoseconds.  The following section will argue that the desired
fluctuations may be found in photon bunching.

\section{Photon bunching}\label{sec:hbt}

Photon bunching, or the HBT effect \cite[named after the pioneering
  experiments of][]{1956Natur.177.27H}, is a quantum-optical
phenomenon, but can be studied semi-classically by first considering a
complex wave and then interpreting the intensity of the wave as
proportional to the probability of detecting photons.\footnote{This
  was first shown by \cite{1963PhRvL..10..277S} and is nowadays called
  the optical equivalence theorem.}

Accordingly, let us consider the wave (a component of the
electromagnetic field) for starlight in a narrow band $S(\nu)$.
The complex wave will then be
\begin{equation}
E(t) = \int e^{2\pi i\nu t} \, S(\nu) \,d\nu \,,
\label{eqn:field}
\end{equation}
and the corresponding intensity will be
\begin{equation}
I(t) \propto \left|E(t)\right|^2 \,.
\end{equation}
If $S(\nu)$ is a delta function, $E(t)$ will just revolve in the
complex plane at a constant rate and $I(t)$ will be a constant.  But
if $S(\nu)$ has a finite width, $E(t)$ will a superposition of
contributions with random phases in the complex plane, making it like
the endpoint of a random walk.  Thus the probability distribution for
$E(t)$ will be Gaussian in the complex plane, and the intensity will
have an exponential distribution.  The larger the frequency spread in
$S(\nu)$ the more quickly $E(t)$ and $I(t)$ will sample their
respective probability distributions.  In particular, if $S(\nu)$
is a Lorentzian
\begin{equation}
S(\nu) \propto \frac1{1 + (2\pi\Delta\tau(\nu-\nu_0))^2}
\label{eqn:spectrum}
\end{equation}
the time scale for the field and intensity to change will be
$\Delta\tau$. This is known as the coherence time.\footnote{The
  precise definition of coherence time varies in the literature. This
  work follows $\Delta\tau_2$ from \cite{1962PPS....80..894M}.}  For a
nanometre filter, the coherence time is roughly one picosecond.  The
intensity auto-correlation
\begin{equation}
1 + g(\tau) \equiv \frac{\big\langle I(t)\, I(t+\tau) \big\rangle}
                   {\big\langle I(t) \big\rangle^2}
\label{eqn:corr}
\end{equation}
will have $g(0)=2$ falling over a time $\Delta\tau$ to $g(\tau)=1$.
This is the standard HBT effect.

Next we consider the effect of magnification by lensing.  For an image
magnified by say $A_1$, the intensity $I(t)$ will get multiplied by
$A_1$.  The complex wave $E(t)$ must get multiplied by $\sqrt{A_1}$
times a phase factor.  The multiplication by $\sqrt{A_1}$ seems a
little mysterious, but can be understood as follows.  Recall that
lensing magnification is the result of light rays, that would not have
reached the observer without lensing, being deflected by the lens
towards the observer.  The corresponding complex waves add, but since
they have random phases they add in random-walk fashion, which on
average produces a factor of $\sqrt{A_1}$.

Now consider a superposition of the waves from two lensed images. The
intensity will be
\begin{equation}
I(t) = \left|\sqrt{\amp_1}E(t) + 
             \sqrt{\amp_2}\,E(t+\Delta\tlens)\right|^2
\label{eqn:intensity}
\end{equation}
The waves $E(t)$ and $E(t+\Delta\tlens)$ are uncorrelated in phase, so
there is no interference.  The intensities, however, may be
correlated.  The auto-correlation (\ref{eqn:corr}) are expected to
show secondary peaks of height $A_1A_2/(A_1+A_2)^2$, or
\begin{equation} \label{eqn:sideband}
g(\pm\Delta\tlens) = \frac{\amp^2-1}{4\amp^2}
\end{equation}
in addition to the main peak of $g(0)=1$.

\begin{figure}
\centering
\includegraphics[width=\hsize]{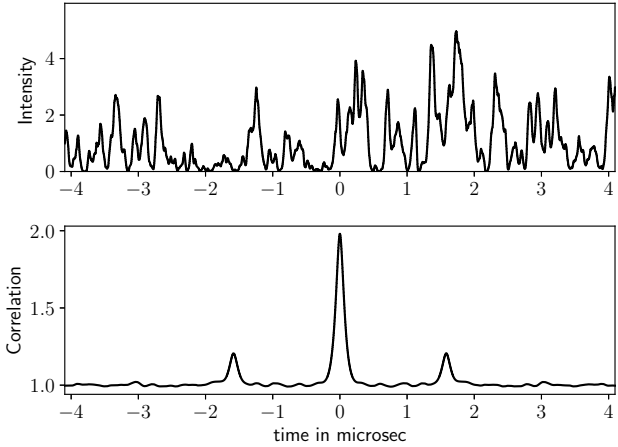}
\caption{\label{fig:signal} A simulation of lensed intensity $I(t)$
  and auto-correlation $1+g(\tau)$ from the system shown in
  Figure~\ref{fig:zones}.}
\end{figure}

Figure~\ref{fig:signal} shows a simulation of lensed intensity and
auto-correlation, obtained as follows.
\begin{enumerate}
\item Each of the 100 zones in Figure~\ref{fig:zones} is taken to emit
  a wave $E(t)$ having a Lorentzian spectral profile with
  $\Delta=10^{-8}\rm\,s$.  The central frequency $\nu_0$ makes no
  difference to the intensity and is set to zero, but the wave from
  each zone is given a random initial phase.  The simulated time
  duration is a millisecond, of which only $\simeq8$~microseconds is
  shown in the figure.
\item Each wave is lensed in the same lensing configuration as in
  Figure~\ref{fig:zones}.  That is, each wave is delayed by the
  appropriate amount and multiplied by $\sqrt{A_1}$ or $\sqrt{A_2}$.
\item All the waves are added, and the intensity computed.
\item Finally the intensity is auto-correlated.  Expected are a
  central peak rising to 2, and (in the assumed lensing configuration)
  secondary peaks at $\pm1.6\rm\,microsec$, each rising to 1.2.
\end{enumerate}
Note that there is no shot noise in Figure~\ref{fig:signal}.  The
intensity fluctuations are just a property of narrow-band light.  They
are well-known in radio astronomy as wave noise
\citep{1999ASPC..180..671R}.  Photon counts will follow a Poisson
distribution with the intensity varying over a time-scale of
$\Delta\tau$.  The result is photon bunching.  The term `super-Poisson
noise' is also used.

The secondary peaks in the auto-correlation are very interesting ---
but are the arguments and simulation valid?  There are at least three
possible concerns.
\begin{itemize}
\item First, it is surprising that the intensity auto-correlation
  appears not to depend on $\nu_0$, only on $\Delta\nu$.
\item Second, there is the assumption that the source can be expressed
  as the sum of small zones.  The simulation is not sensitive to the
  number of zones.  On the other hand, the sample points representing
  zones are $\sim0.5$~light-sec part, which is an order of magnitude
  larger than the Fresnel-zone length $\sqrt{\lambda D_S}$ associated
  with diffraction.  Actually, diffraction effects are possible on
  even smaller scales, though only near lensing caustics
  \citep[e.g.,][]{2003ApJ...594..456Z} which does not apply to most
  microlensing events.
\item Third, the factors of $\sqrt{A_1}$ and $\sqrt{A_2}$ in the total
  intensity (\ref{eqn:intensity}) would come with arbitrary phases;
  but again, these would not destroy the auto-correlation.
\end{itemize}
A laboratory experiment to test for time-delayed photon bunching is
desirable.  A suitable variant of the experiments by
\cite{2015A&A...580A..99D} and \cite{2016MNRAS.457.4291T} could do
such a test.

\section{Signal-to-noise considerations}\label{sec:snr}

Assuming time-delayed photon bunching is real, identifying the most
promising events in advance would not be a problem.  Although Galactic
microlensing events are rare, with at most a few stars per million
being lensed at any given time \citep{2013ApJ...778..150S}, an
early-warning system \citep[such as in][]{2015AcA....65....1U} would
provide the expected image brightness and magnification.  The
challenge would be getting sufficient signal-to-noise to measure the
time-delayed photon bunching on even the best candidates.

The basic setup and signal-to-noise considerations would be similar to
those in recent work on intensity interferometry
\citep{2017MNRAS.467.3048P,2018SPIE10701E..21L,2018SPIE10701E..0XW}.
The starlight is filtered to a very narrow wavelength band, to
increase the coherence time $\Delta\tau$, and then photons are counted
with a time-resolution $\Delta t$. The favoured detector technology is
single-photon avalanche photodiodes.  Multiple photodiodes, each
devoted to one narrow wavelength range, can be used in parallel, thus
having multiple photon-counting channels.

In intensity interferometry there are two (or more)
telescopes whose photon counts are cross-correlated, whereas for
microlensing only one telescope is required --- but a bigger one,
because the targets are fainter.  An additional requirement in
microlensing is to have the coherence time $\Delta\tau$ comparable to
or longer than the time-delay spread $\Delta\Delta\tlens$.  The latter
condition could be checked in advance, since an approximate source
size $\Delta\beta$ would be available for an ongoing microlensing
event, providing ballpark estimates for $\Delta\tlens$ and
$\Delta\Delta\tlens$.

The SNR can be estimated by adapting an argument from intensity
interferometry, as follows.  Let $r$ be the rate of photons arriving
per unit collecting area in some narrow spectral band, and let
$\Delta\tau\sim1/\Delta\nu$ be the coherence time corresponding to
that spectral band.  If the telescope has unit collecting area and
perfect detection efficiency, the number of photons per coherence time
will be $r\,\Delta\tau$.  Now consider two time bins, each of duration
$\Delta t$ (the instrumental time resolution), but separated by
$\Delta\tlens$.
\begin{itemize}
\item Each time bin will contain $\Delta t/\Delta\tau$ time slices,
  during which the light is coherent.  In a pair of coherent time
  slices $\Delta\tlens$ apart, there will be
  $g(\Delta\tlens)\times(r\,\Delta\tau)^2$ pairs of HBT-correlated
  photons.  Hence, there will be $g(\Delta\tlens)\times
  r^2\Delta\tau\,\Delta t$ HBT-correlation events per time bin.  This
  is the signal.
\item Meanwhile in the same time bins, there will $(r\,\Delta t)^2$
  pairs of photons correlated by chance.  This number is the
  background, and $r\,\Delta t$ is the corresponding noise.
\end{itemize}
The SNR per time bin $\Delta t$ is thus $g(\Delta\tlens)\times
r\,\Delta\tau$.  This applies to unit collecting area and perfect
detectors.  If we have collecting area $A$ and photon-detection
efficiency $\gamma$, these factors just multiply $r$.  Concerning the
spectral bandpass, narrowing it reduces $r$ but increases $\Delta\tau$
by the same factor, and (remarkably) leaves the SNR per time bin
unaffected.  Hence, it is advantageous to have many narrow spectral
channels.  For $N$ channels, the SNR gets multiplied by $\sqrt N$.
Similarly, if the total observing time is $T$, the SNR will be
multiplied by $\sqrt{T/\Delta t}$.  The result is
\begin{equation} \label{eqn:snr}
\hbox{SNR} \sim g(\Delta\tlens) \times \gamma A \, r\,\Delta\tau
                \left(\frac{NT}{\Delta t}\right)^{1/2}
\end{equation}
provided $\Delta t \gtrsim \Delta\tau$.  The SNR improves as the time
resolution gets smaller, until it becomes comparable to the coherence
time. Lowering $\Delta t$ still further does not help, as
super-Poisson noise takes over \citep{2014MNRAS.437..798M}.  The
factor $g(\Delta\tlens)$ will be less than one, but roughly
compensating for that is the increase in $r$ from lensing
amplification.  So it is reasonable to consider $(\gamma
A)\times(r\,\Delta\tau)\times\sqrt{NT/\Delta t}$ without lensing.

To reach a reasonable SNR in one night, one would need to achieve
$\hbox{SNR}\sim1$ in $T=10^3{\rm s}$.  If $N=10$ channels each with
$\Delta t=10^{-8}{\rm s}$ are installed, $\sqrt{NT/\Delta t}\sim10^6$.
Hence one needs $(\gamma A)\times(r\,\Delta\tau) \sim 10^{-6}$ at
least.  The next generation of extremely large telescopes and highly
efficient detectors could offer $\gamma A\sim 10^3\rm\,m^2$ at best.
This suggests that sources down to
$r\,\Delta\tau\sim10^{-9}\rm\,m^{-2}$ would be plausible targets.

\begin{figure}
\centering
\includegraphics[width=\hsize]{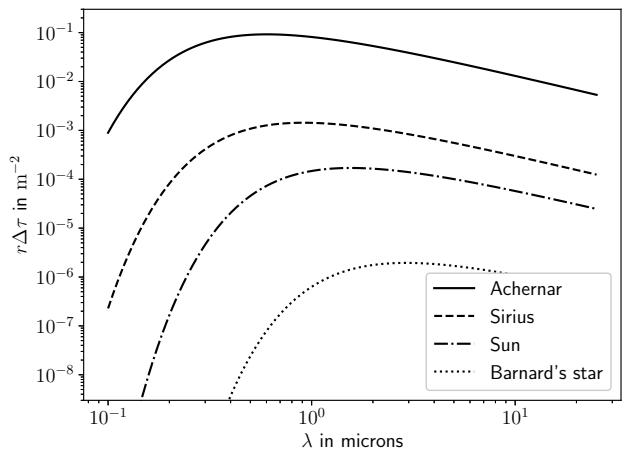}
\caption{Curves of $r\,\Delta\tau$ for four different values of radius
  and surface temperature (corresponding approximately to the named
  stars) but viewed from 1~parsec in all cases.}
\label{fig:flux}
\end{figure}

The photon flux per coherence time $r\,\Delta\tau$ is simply the
spectral flux density divided by energy and integrated over the
source.  Figure \ref{fig:flux} shows some example curves of
$r\,\Delta\tau$.  A blackbody disc of radius $R_\odot$ and
$T=5800\,\rm K$, viewed from 1~pc is labelled ``Sun''.  A disc of
radius $0.2R_\odot$ and $T=3100\,\rm K$ viewed from the same distance
is labelled ``Barnard's star''.  Similarly, the label ``Sirius''
corresponds to $1.7R_\odot$ and $9900\,\rm K$ from 1~pc, and
``Achernar'' to $9R_\odot$ and $T=15000\,\rm K$ from 1~pc.  From such
a short distance, any of these would be a plausible candidate.  But if
they are located at $\sim10\rm\,kpc$, the value of $r\,\Delta\tau$
falls by a factor of $10^8$.  This leaves only ``Achernar'' as a
plausible candidate.

The above suggests that photon bunching would be measurable in O or B
main-sequence stars at $\sim10\rm\,kpc$.  (Giant or supergiant stars
of similar brightness are unlikely to be useful, because $\Delta\beta$
will become comparable to $\thee$, washing out the time-delay peaks.)
Microlensed early-type stars must be exceedingly rare, and it is not
clear that any have been observed yet.  \cite{2009AcA....59..255N}
show some candidate light curves in their Figure~5, which may have
been such, if they were indeed microlensing events.

Plans for intensity interferometers include proposals to create light
buckets as large as $10^4\rm\,m^2$ by attaching additional detectors
to air Cerenkov telescopes \citep{2016SPIE.9907E..0MD}.  The mirrors
involved are not optical quality (hence the huge areas).  These
sacrifice image quality but remain adequate for collecting light from
bright stars.  Microlensing surveys, however, need to target crowded
fields of stars, and to count photons from a single star in a crowded
field, good image quality is essential.  Hence a general-purpose
telescope cannot be substituted, and there is no way of increasing $A$
in Eq.~(\ref{eqn:snr}) in the near future.  Any hope for increased
sensitivity would lie in increasing the number of spectral channels
$N$.  Small arrays of single-photon avalanche photodiodes have been
tested \citep[e.g.,][]{2014SPIE.9114E..0CT}, and perhaps much larger
arrays may be possible in the future.  Another class of detectors with
prospects for many-channel photon counting are superconducting
nanowires \citep[e.g.,][]{2015OExpr..2333792V}.

\def\aap{A\&A}
\def\araa{ARA\&A}
\def\actaa{Acta Astr}
\def\apj{ApJ}
\def\apjl{ApJL}
\def\mnras{MNRAS}
\def\nat{Nature}
\def\pasj{PASJ}
\def\procspie{Proc.\ SPIE}

\bibliographystyle{mn2e}
\bibliography{ms.bbl}

\end{document}